%% file: paper.tex
\documentclass[aps,physrev,reprint,superscriptaddress,amsmath,amssymb]{revtex4-2}

\usepackage{anyfontsize}
\usepackage{braket}
\usepackage[font=small]{caption}
\usepackage{indentfirst}
\usepackage{calc}
\usepackage{xcolor}
\usepackage{graphicx}
\usepackage{subcaption}
\usepackage{multirow}
\usepackage{dcolumn}
\usepackage{bm}
\usepackage{ifthen}
\usepackage{upgreek}
\usepackage{svg}
\usepackage{soul}
\usepackage{textgreek}
\usepackage{upgreek}
\usepackage[hidelinks,colorlinks]{hyperref}

\soulregister\ref7
\soulregister\eqref7
\soulregister\autoref7
\soulregister\nameref7
\soulregister\cite7
\soulregister\textcite7

\hypersetup{
    citecolor=blue,
    linkcolor=blue,
    urlcolor=blue,
    filecolor=black,
    menucolor=black
}

\renewcommand\vec{\mathbf}

\begin{document}

\title{Single-photon loading of polar molecules into an optical trap}

\author{Bart J. Schellenberg}\affiliation{Van Swinderen Institute for Particle Physics and Gravity, University of Groningen, Groningen, The Netherlands}\affiliation{Nikhef, National Institute for Subatomic Physics, Amsterdam, The Netherlands}
\author{Eifion H. Prinsen}\affiliation{Van Swinderen Institute for Particle Physics and Gravity, University of Groningen, Groningen, The Netherlands}\affiliation{Nikhef, National Institute for Subatomic Physics, Amsterdam, The Netherlands}
\author{Janko Nauta}\affiliation{Experimental Physics Department, CERN, Geneva, Switzerland}
\author{Lukáš F. Pašteka}\affiliation{Van Swinderen Institute for Particle Physics and Gravity, University of Groningen, Groningen, The Netherlands}\affiliation{Nikhef, National Institute for Subatomic Physics, Amsterdam, The Netherlands}\affiliation{Department of Physical and Theoretical Chemistry, Comenius University, Bratislava, Slovakia}
\author{Anastasia Borschevsky}\affiliation{Van Swinderen Institute for Particle Physics and Gravity, University of Groningen, Groningen, The Netherlands}\affiliation{Nikhef, National Institute for Subatomic Physics, Amsterdam, The Netherlands}
\author{Steven Hoekstra}\email{s.hoekstra@rug.nl}\affiliation{Van Swinderen Institute for Particle Physics and Gravity, University of Groningen, Groningen, The Netherlands}\affiliation{Nikhef, National Institute for Subatomic Physics, Amsterdam, The Netherlands}

\date{\today}

\begin{abstract}
We propose a scheme to transfer molecules from a slow beam into an optical trap using only a single photon absorption and emission cycle. The efficiency of such a scheme is numerically explored for BaF using realistic experimental parameters. The technique makes use of the state-dependent potential in an external electric field to trap molecules from an initial velocity of order $10\,\text{m}/\text{s}$. A rapid optical transition at the point where the molecules come to a standstill in the electric field potential irreversibly transfers them into a $\sim7\,\text{mK}$ optical lattice trap. For a pulsed Stark decelerated beam, we estimated the per-shot efficiency to be $\sim0.52\,\%$ or up to $\sim10^4\,\text{molecules}$, with a potential factor $2$ improvement when the fields are synchronously modulated with the arriving velocity components. The irreversibility of the scheme allows for larger numbers to be built up over time. Since this scheme does not rely on a closed cycling transition for laser cooling, it broadens the range of molecules that can be used for research on cold molecular chemistry, quantum information, and fundamental interactions in optical traps.
\end{abstract}
\maketitle
\section{Introduction}
\indent Trapped molecules are versatile systems that enable studies in many fields~\cite{carr_2009} such as cold molecular chemistry~\cite{krems_2008,ospelkaus_2010,Stuhl_2014,Zhao_2022,liu_2022,karman_2024}, quantum information processing~\cite{demille_2002,cornish_2024}, and fundamental interactions~\cite{acme_2018,roussy_2023}. An optical trap is particularly attractive as it can trap molecules in states independent of their Stark or Zeeman shifts. Especially when operated at so-called magic conditions, where the AC-Stark shifts of two levels of interest are tuned to be the same, it offers an environment highly suitable for precision measurements. Although successful laser cooling~\cite{shuman_2010,hummon_2013,zhelyazkova_2014,kozyryev_2017,truppe_2017,lim_2018,Anderegg_2018,rees_2020,mitra_2020,rockenhauser_2024,dai_2024} and magneto-optical trapping~\cite{hummon_2013,anderegg_2017,truppe_2017,collopy_2018,baum_2020,vilas_2022,burau_2023,zeng_2024,li_2024,barry_2024} have been demonstrated for a number of diatomic and polyatomic species, further cooling and compression is required to transfer molecules into an optical trap. In particular polyatomic molecules, consisting of three or more atoms, offer additional opportunities compared to diatomic systems due to their rich internal structures~\cite{wall_2015,augustovicova_2019,hutzler_2020,bause_2024}. The same complexity of these energy levels also brings increased challenges when it comes to cooling. This primarily stems from the need to scatter a large number of photons without significant losses from the optical cycle, as well as the ability to sufficiently compress the molecule cloud. Consequently, alternative routes that do not suffer from the limitations introduced by laser cooling have been studied on a number of occasions~\cite{narevicius_2009,lu_2014,singh_2023,amit_2025}.\\
\begin{figure}[ht]
    \centering
    \vspace*{-.5cm}
    \resizebox{.48\textwidth}{!}{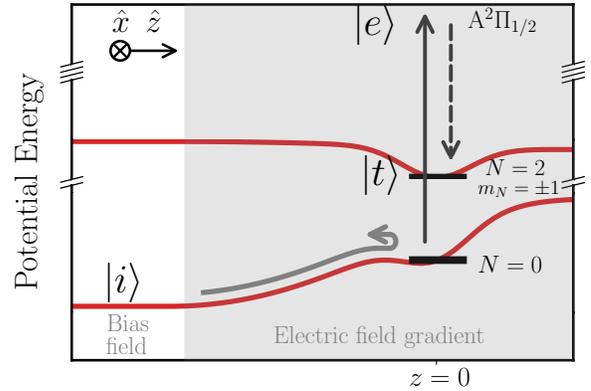}
    \captionsetup{width=.45\textwidth}
    \caption{\textit{A schematic of the single-photon loading scheme. Molecules approach from the left whilst in some initial state $\ket{i}$ and fly along the $+\hat{z}$ axis. At their turning point in the potential, they are pumped to an excited state $\ket{e}$, from which there is a chance to decay to the trapped state $\ket{t}$. The center of the optical trap is located at $z=0$. A fraction of the molecules remain trapped in this state by the optical trap.}}\label{figScheme}
\end{figure}
\indent In this work we propose a single-photon loading scheme with the purpose of directly transferring molecules from a slow beam into an optical trap. The scheme combines the state-dependent energy landscape in an externally applied electric field with an irreversible electronic transfer at a point where the kinetic energy of the molecule is minimized. Similar experiments using magnetic fields were previously designed~\cite{price_2007,narevicius_2009,jayich_2014} and performed~\cite{price_2008,riedel_2011,lu_2014} for both atoms and molecules. We numerically explore the efficiency of this process for the BaF molecule, as a prototype molecule that is interesting for precision experiments~\cite{aggarwal_2018,li_2023,boeschoten_2024,corriveau_2024}.\\
\indent The general idea is illustrated in Figure~\ref{figScheme}. A more detailed scheme with suitable states and parameters specific for $\text{BaF}$ is worked out in section~\ref{secParameters}. Generally, we start by preparing the molecules in some long-lived initial state $\ket{i}$, whose Stark curve strongly depends on the electric field. The molecules initially fly along the $+\hat{z}$ direction into an electric field gradient, which forms a repulsive potential. At the classical turning point inside the potential, where the molecules have minimal kinetic energy, we place an optical trap. Once the molecules reach this optical trap, we electronically excite them to some short-lived excited state $\ket{e}$, after which they promptly decay. If the molecules decay to a state whose Stark shift stays approximately constant within the same electric field, they will remain inside the optical trap. We call this final state the trapped state $\ket{t}$. The intermediate state $\ket{e}$ is chosen such that a laser driving the $\ket{i}\to\ket{e}$ transition does not also resonate with $\ket{t}$, turning the scheme into a one-way barrier, or "molecule diode". Previous work discussed a similar scheme as a one-dimensional realization of Maxwell's demon~\cite{bannerman_2009}.\\
\indent The lowest rotational states of heteronuclear diatomic and linear polyatomic molecules tend to feature significant high-field-seeking (HFS), low-field-seeking (LFS), as well as in-between states whose Stark shift remain approximately constant~\cite{meerakker_2012}. The Stark shifts associated with these states correspond to maximum stopping velocities up to a few tens of meters per second, ensuring a fully adiabatic evolution into experimentally achievable field gradients. In more general polyatomic molecules, where the rotational manifold becomes larger, several decay channels from the excited state $\ket{e}$ may end in states suitable for $\ket{t}$. Since our only requirement for $\ket{t}$ is an approximately constant Stark curve, and we do not rely on populating a single specific rotational level for loading, each of these decay channels may contribute to the overall loading efficiency. In both HFS and LFS cases, the $\ket{i}$ state molecules are prepared inside some weak bias electric field. This ensures that the projection states represent good quantum numbers, whilst keeping state mixing following the Stark effect at larger fields at a minimum. In the case where $\ket{i}$ is HFS, the bias is then temporally increased such that the molecules are able to proceed into a spatially decreasing field used for slowing.
\section{Molecular Interactions}
\begin{figure}[ht]
    \centering
    \resizebox{.48\textwidth}{!}{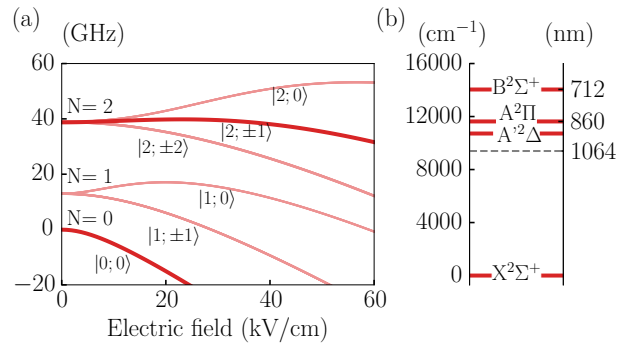}
    \captionsetup{width=.45\textwidth}
    {\phantomsubcaption\label{figEnergyStark}}
    {\phantomsubcaption\label{figEnergyBig}}
    \caption{\textit{a) The Stark curves of the lowest rotational levels (notation $\ket{N;m_N}$) in the $X^2\Sigma^+\;v=0$ vibronic state of $\text{BaF}$; b) The lowest electronic states in $\text{BaF}$. The dashed line indicates the wavelength used for the optical trap.}}
\end{figure}
\indent The Stark curves for the lower lying rotational states of the electronic ground state of BaF are shown in Figure~\ref{figEnergyStark}. These curves were calculated by PGopher~\cite{pgopher_2017} using molecular constants from Ryslewicz \textit{et al.}~\cite{ryzlewicz_1980} and Ernst \textit{et al.}~\cite{ernst_1986}. In this analysis we neglect the hyperfine splitting, as this concerns energy shifts much smaller than the rotational shift ($\sim1:100$) inside strong electric fields ($\sim10\,\text{kV}/\text{cm}$). The interaction of the molecules with the external electrostatic field is described by $H^{\text{Stark}}=-\vec\mu_e\cdot\vec{E}$ (with $\mu_e$ being the electric dipole moment of BaF and $\vec{E}$ the external field), and its interaction with the optical field by $H^{\text{opt}}=-\vec{d}_e\cdot\vec{\mathcal{E}}$ (where $\vec{d}_e$ represents the dipole operator and $\vec{\mathcal{E}}$ the electric component of the optical field.) When $H^{\text{Stark}}$ remains sufficiently small compared to the rotational splitting, $H^{\text{opt}}$ can be expanded into spherical tensor components~\cite{kien_2013,caldwell_2020}
\begin{align*}
    H^{\text{opt}}_{ij}&=-\frac{\mathcal{E}_0^2}4\sum\limits_{k=0}^2\mathcal{A}^k[\alpha(\omega)]\cdot \mathcal{P}^k(\hat{\vec{\mathcal{E}}},\hat{\vec{\mathcal{E}}}^*),\\
    &=-\frac{\mathcal{E}_0^2}4\bigg(\delta_{ij}\alpha^{\text{S}}(\hat{\mathcal{E}}\cdot\hat{\mathcal{E}}^*)+\addtocounter{equation}{1}\tag\theequation\label{eqHOPT}\\
    &\;\;\;\;\;\;+\epsilon_{ijk}\alpha^{\text{V}}_{k}(\hat{\vec{\mathcal{E}}}\times\hat{\vec{\mathcal{E}}}^*)_k+\alpha^{\text{T}}_{ij}(\hat{\vec{\mathcal{E}}}\otimes\hat{\vec{\mathcal{E}}}^*)_{ij}\bigg),
\end{align*}
where $\mathcal{A}^k$ and $\mathcal{P}^k$ are spherical tensors of rank $k$, taking into account the molecular polarizability and optical polarization respectively. The contributions of the scalar ($\alpha^{\text{S}}$), vector ($\alpha^{\text{V}}$), and tensor ($\alpha^{\text{T}}$) components of the polarizability have been made explicit with the second equality. Summation over $k$ is assumed, $\delta_{ij}$ represents the Kronecker delta, $\epsilon_{ijk}$ is the Levi-Civita symbol, and $\alpha^{\text{T}}_{ij}$ is a symmetric traceless operator. When the light is linearly polarized, the unit vector $\hat{\vec{\mathcal{E}}}$ is real and the vector term vanishes. The scalar component is isotropic and consequently independent of the molecular projection states. The tensor part is state-dependent and can introduce deviations from isotropy at the order of $10\%$. For an explicit derivation of the relevant operators, the reader is referred to the work by Caldwell \textit{et al.}~\cite{caldwell_2020}. We found that whilst the introduction of the tensor polarizability can affect the trap depth at the order of $10\%$, it does not significantly impact the loading efficiency. This is because the molecules that are successfully transferred into the trap predominantly possess energies near the trapping threshold, which follows from the inevitable gain in momentum due to the attractive forces of the optical trap.\\
\renewcommand{\arraystretch}{1.2}
\begin{table}[!hbpt] 
    \vspace*{-.5cm}
    \centering
    \begin{tabular}{ccccc}
        &&$\omega=0$&$\omega=c/1064\,\text{nm}$\\
        \cline{1-4}
        $\alpha_\parallel\;\;\;\;$&X$^2\Sigma$&$169.82(77)$&$288$\\
        &A$^2\Pi_{1/2}$&$137.4(5.2)$&\multirow{2}{*}[0.5ex]{$-27$}\\
        &A$^2\Pi_{3/2}$&$136.3(3.1)$&\\
        \cline{1-4}
        &&&\\
        \cline{1-4} 
        $\alpha_\perp\;\;\;\;$&X$^2\Sigma$&$277.8(1.4)$&$671$\\
        &A$^2\Pi_{1/2}$&$-1275(184)$&\multirow{2}{*}[0.5ex]{$-382$}\\
        &A$^2\Pi_{3/2}$&$-910(104)$&\\
        \cline{1-4}
    \end{tabular}
    \caption{Parallel ($\alpha_\parallel$) and perpendicular ($\alpha_\perp$) static and dynamic polarizabilities of BaF in atomic units. At $\omega=c/1064\,\text{nm}$, the components for X$^2\Sigma$ differ from $\omega=0$ primarily through coupling to the A$^2\Pi$ ($\alpha_\perp$) and B$^2\Sigma$ ($\alpha_\parallel$) states, and for A$^2\Pi$ by X$^2\Sigma$ ($\alpha_\perp$) and C$^2\Pi$ ($\alpha_\parallel$).}
    \label{tab:alpha}
\end{table}
\renewcommand{\arraystretch}{1.0}
\;\\
\indent To determine the optical trap depth and the corresponding optical intensity that is required, we have calculated the polarizability of the molecule in the involved electronic states. The static polarizability of BaF in the X$^2\Sigma$, A$^2\Pi_{1/2}$, and A$^2\Pi_{3/2}$ states was calculated using the finite-field procedure based on the energies obtained by the DIRAC23 program package~\cite{dirac-paper,DIRAC23} using the Fock-space coupled cluster method~\cite{Kaldor1991} and an exact 2-component Hamiltonian~\cite{ilias2007}. Augmented valence Dyall basis sets~\cite{Dyall2009,Dyall2016} were used in the calculations. Cardinality and augmentation were increased until further refinements improved the polarizability at the order of only a few percent. The final results were calculated using the t-aug-dyall.v4z basis set. Similarly, various other computational parameters (level of theory, coupled cluster active space, relativistic Hamiltonians) were investigated. Based on this study, we assigned the theoretical uncertainties to the calculated values.\\
\indent The dynamic polarizability was calculated at the coupled cluster (CCSD) level using the linear response method as implemented in the CFOUR program package~\cite{cfour-paper,cfour}. Relativistic effects were incorporated by use of Dunning's pseudopotential basis sets~\cite{Kendall1992,Lim2006,Hill2017}, with an effective core potential on the barium atom, replacing its 46 innermost electrons. The final results were calculated using the aug-cc-pV5Z-PP basis set. Note that this scalar relativistic scheme does not account for spin-orbit splitting, and thus only yields the polarizability for the $A^2\Pi$ state as a whole. A summary of the relevant quantities is given in Table \ref{tab:alpha}. The details of the computational study are presented in the upcoming theoretical paper~\cite{Prinsen2025}.\\
\section{Parameters}\label{secParameters}
\indent Within the context of this work we have selected to use a $\lambda=1064\,\text{nm}$ trapping laser, as this wavelength is well red-detuned from the lowest electronic transition and intense lasers with good intensity and frequency stability are readily available at this wavelength. We will consider an optical lattice trap formed by the standing wave light field inside an enhancement cavity. The use of an optical cavity offers a number of experimental advantages: it uses the incident laser power more efficiently, enabling the formation of a larger trap volume. Larger trap volumes reduce collisional effects when considering large molecule numbers, and they also increase the geometric overlap with an incoming source beam. Moreover, a wide optical trap allows a smoother deformation in the $\ket{i}$ state potential, which reduces the relative height of the local maximum in the potential for incoming molecules. In addition, by implementing a polarization filter within the cavity, a high degree of polarization purity can be obtained~\cite{zhu_2013}. In a ring type optical cavity, molecules can be directly transported to an environment suitable for precision measurements, using an optical conveyor belt~\cite{bause_2024}.\\
\indent When the separation of the cavity mirrors is large compared to the beam waist $w_0$, the optical field of the trap reduces to that of a Gaussian standing wave. The resulting 1D lattice trap has an amplitude $\mathcal{E}_0$ that can be obtained from the incident laser power $P$ and the cavity finesse $\mathcal{F}$ through
\begin{gather}
    \mathcal{E}_0=\sqrt{\frac{4P\mathcal{F}}{\pi^2w_0^2c\epsilon_0}}\label{eqCavityE0},
\end{gather}
which follows from integrating its transverse profile. In the absence of the external electric field, using equation \eqref{eqCavityE0} with \eqref{eqHOPT} allows us to directly connect the trap depth with the configurable parameters $P$, $w_0$, and $\mathcal{F}$. For instance, by coupling $P=20\,\text{W}$ into a low finesse cavity of $\mathcal{F}=500$ and $w_0=150\,\text{μm}$, a $10\,\text{mK}$ deep trap can be formed. The trap is oriented along the $\hat{x}$ direction, which lies orthogonal to the molecular beam. Its potential takes the form of many closely-spaced discs, sometimes also referred to as 'pancakes'. Each disc has a size of $dx\sim0.5\,\text{μm}$ and $dy,dz\sim100\,\text{μm}$. It is to be noted that in the presence of the electric field, the combination of the AC and DC Stark shift skews the potential, which slightly reduces the actual depth. For our example, we numerically computed an effective trap depth of about $7.3\,\text{mK}$ for the above-mentioned configuration.\\
\begin{figure*}
    \centering
    \resizebox{.86\textwidth}{!}{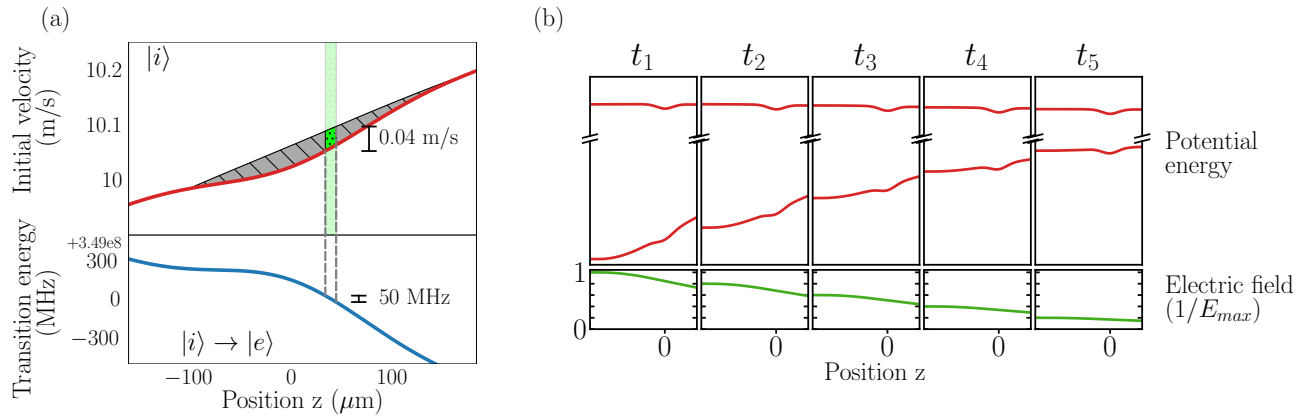}
    \captionsetup{width=.9\textwidth}
    {\phantomsubcaption\label{figAcceptance}}
    {\phantomsubcaption\label{figSweep}}
    \vspace*{.7cm}
    \caption{\textit{a) Top: phase space acceptance of molecules arriving in $\ket{i}$ into the trapped state $\ket{t}$. The solid red curve represents the potential following the AC and DC Stark effects, and effectively shows the position $z$ where molecules with a given initial velocity will come to a standstill. Bottom: position-dependent transition energy for $\ket{i}\to\ket{e}$. Further details are elaborated in the main text; b) (Not to scale.) Modulating the external fields with time allows different velocity components to be captured, as the relative height of the trap in $\ket{i}$ decreases.}}
\end{figure*}
\indent To optimize the shape of the electric potential for the loading scheme, we consider the difference between the electrostatic field at the position of the optical trap and the upstream electric field to determine the height of the initial barrier in $\ket{i}$. We also optimize the slope of the electric field at the trap, which determines the trap depth in $\ket{i}$. For an idealized configuration, a steep slope is preferred as this prevents the molecules from regaining momentum when they enter the optical field. At the same time, the height of the initial barrier in $\ket{i}$ is tuned to match the initial energy of the incoming molecules, such that they come to a standstill close to the optical trap.\\
\indent The maximum capture velocity is determined by the shape of the Stark curve. For instance, by using an increasing electric field with $+\hat{z}$, molecules from the LFS $\ket{i}=\ket{1;0}$ state can be brought to a stand-still if their initial velocity was no greater than $4.5\,\text{m}/\text{s}$, before they reach their HFS regime beyond $19\,\text{kV}/\text{cm}$. Alternatively, molecules in the $\ket{i}=\ket{2;0}$ state from up to $8.5\,\text{m}/\text{s}$ can be stopped, but at the cost of requiring a much larger electric field ($56\,\text{kV}/\text{cm}$). In order to capture molecules from a larger initial velocity using a more readily achievable electric field, whilst also providing a suitable candidate for $\ket{t}$, we found it better to reverse the situation. That is, one can use a decreasing electric field with $+\hat{z}$ to slow down molecules in a HFS state. Whilst the transverse stability for HFS states is a point of attention, deceleration under similar conditions has been achieved before~\cite{bethlem_2006,wohlfart_2008}. A loading scheme involving HFSs can be achieved using $\ket{i}=\ket{0;0}$ and $\ket{t}=\ket{2;\pm1}$ (see also Figure~\ref{figEnergyStark}). Starting at $40\,\text{kV}/\text{cm}$, these molecules can be stopped from up to $14\,\text{m}/\text{s}$. In the same electric field range, the $\ket{t}=\ket{2;\pm1}$ level remains mostly unaffected. Since both states have a positive parity, and the transitions happen in a relatively weak electric field, which reduces state-mixing, they can be directly coupled using an intermediate state $\ket{e}$ of negative parity.
\section{Loading}
\indent Having identified the most important parameters of the loading scheme, let us discuss the efficiency with which incoming molecules can be transferred into the optical trap. The above-mentioned trap depth of $7.3\,\text{mK}$ corresponds to at most $0.88\,\text{m}/\text{s}$ in kinetic energy, whilst the typical timescales involved in the electronic transitions range between tens to hundreds of nanoseconds~\cite{hao_2019,aggarwal_2019}. This means that during the optical transitions, the molecule does not move significantly inside the micrometer-sized trap, and we may consider the transfer $\ket{i}\to\ket{e}$ and subsequent decay $\ket{e}\to\ket{t}$ as an instantaneous process. This allows us to determine for each point in position-velocity phase space whether an instantaneous transfer at that point can move the molecule to inside the final optical potential of state $\ket{t}$. The corresponding phase space along $(z,v_z)$ is shown in the upper half of Figure~\ref{figAcceptance}, indicated by the gray shaded (striped) areas.\\
\indent The next step is to consider the conditions under which a resonant excitation $\ket{i}\to\ket{e}$ can be driven. Since the differential Stark shift in the electric and optical fields depends on the position of the molecule, the linewidth of the excitation can be up to a few hundred $\text{MHz}$, as is indicated by the lower half of Figure~\ref{figAcceptance}. This means that a laser source with a large linewidth is preferred. Using an optical modulator, we intend to broaden the pump light such that molecules within a $50\,\text{MHz}$ linewidth can be excited. The green shaded (dotted) area in Figure~\ref{figAcceptance} illustrates the part of the phase space where the excitation energy lies within a $50\,\text{MHz}$ bandwidth and where the molecules can be successfully transferred for this particular example. At the same time, it is important to ensure that no transitions involving the already trapped molecules in $\ket{t}$ lie within the same bandwidth. For our case, we choose $\ket{e}=\ket{J=3/2;-}$ of the $A^2\Pi_{1/2}\;v=0$ manifold, such that the excitation energy is sufficiently far detuned from any transition involving $\ket{t}$. The lifetime of the $A^2\Pi_{1/2}$ states in BaF is $57.1(3)\,\text{ns}$~\cite{aggarwal_2019}, with about $95\%$ vibrational branching to $X^2\Sigma^+\;v=0$~\cite{chen_2016,hao_2019}. The rotational branching ratios to $\ket{i}=\ket{0;0}$ and $\ket{t}=\ket{2;\pm1}$ in $X^2\Sigma^+$ are $67\%$ and $25\%$ respectively~\cite{chen_2016}. This means that a single absorption-emission cycle has a $24\%$ chance of producing a trapped molecule $\ket{t}$, $63\%$ chance of resetting the molecule back to the initial state $\ket{i}$, and $13\%$ chance to be lost to some alternative state. By rapidly re-exciting the molecules between $\ket{i}$ and $\ket{e}$ until they decay to $\ket{t}$, we can thus transfer at best $\frac{0.24}{1-0.63}\approx65\%$ to $\ket{t}$. After a successful decay to $\ket{t}$, the full entropy is dissipated through a single photon.\\
\indent The efficiency with which molecules can be transferred into the optical trap depends on the matching of the trap and electric field parameters with the properties of the incident molecular beam. As an example, let us consider a Stark decelerated molecular beam~\cite{aggarwal_2018}, whose molecules travel at a longitudinal velocity of $v_z=(10\pm1)\,\text{m}/\text{s}$, within a $dx,dy=1\,\text{mm}$ diameter, and with a near zero transverse velocity. The spatial overlap of such a molecular beam with the thin ($\sim150\,\text{μm}$) optical lattice results in about $4.1\%$ of the molecules traversing a region in the optical field that contains a local minimum in the $\ket{t}$ state potential. The velocity acceptance, as is indicated by the green shaded (dotted) area in Figure~\ref{figAcceptance} for the on-axis case, integrates to $4.1\%\,\times\,1.4\%$ of the molecules also having the right velocity to be captured. By taking into account that at best $65\%$ of these molecules can make it to the $\ket{t}$ state, we estimate to capture $\sim0.04\%$ of the incoming molecules. In the case of a Stark decelerated beam that delivers $\sim2\times10^6$ molecules at $10\,\text{Hz}$\,\cite{aggarwal_2018}, this results in $\sim750$ trapped molecules per shot.\\
\indent The velocity acceptance of the loading scheme forms the primary limitation to the capture rate. For the configuration illustrated by Figure~\ref{figAcceptance}, the velocity acceptance is approximately $10\pm0.02\,\text{m}/\text{s}$. In the above-mentioned example using a Stark decelerated beam, only $1.4\%$ of the molecules fall within the velocity acceptance. To better capture the velocity class produced by the Stark decelerator, we introduce two potential improvements. In both cases, we propose to use a pulsed molecular beam where the molecules are left to propagate freely for some time prior to entering the electric field gradient. The free propagation allows faster molecules to get ahead of the slower ones, temporally separating their arrival time at the trap.\\
\indent The first improvement involves synchronously sweeping the frequency of the pump laser. Here we use the fact that each velocity class has its turning point at a distinct differential Stark shift (see lower plot in Fig~\ref{figAcceptance}), and that these velocity classes are now separated in time. A synchronous sweep of the laser frequency provides an improvement of the velocity acceptance from $1.4\%$ to $20\%$, leading to a total capture efficiency of $0.52\%$, or $10^4$ molecules per shot.\\
\indent A second improvement may be realized by also sweeping the strength of the electric fields, as illustrated in Figure~\ref{figSweep}. We expect the capture rate using a Stark decelerated beam to potentially improve to about $1\%$, which is when the on-axis velocity capture efficiency is increased to one. Further investigation of this dynamic scheme could be done by simulating the molecular trajectories through the setup.
\section{Applicability}
\indent Once the molecules are trapped, they can either be used directly to perform a (precision) measurement, or they can be transported using an optical conveyor belt. The latter can be achieved by designing the optical trap using a ring type cavity, and by coupling two counter propagating laser beams whose frequencies are slightly detuned, effectively moving the trap minima at a velocity that is proportional to the detuning. Performing experiments directly after capturing may be limited due to the presence of the static electric fields, but also photon scatterings due to the off-resonant trapping laser may become problematic at high intensities. In a far-off resonance trap, the scattering rate can be estimated through~\cite{grimm_2000}
\begin{gather}
    \gamma_{\text{scat}}\approx\frac{U}\hbar\frac\Gamma\Delta\label{eqGamma},
\end{gather}
where $\gamma_{\text{scat}}$ represents the photon scattering rate, $U$ the trap depth, $\Gamma$ the linewidth of the transition, and $\Delta$ the detuning of the trapping light with that transition. For our case, we estimate $\gamma_{\text{scat}}=5\,\text{Hz}$. The scattering rate could be reduced by: 1) reducing the trap depth, which in turn increases the velocity sensitivity of the loading scheme. This results in fewer trapped molecules that would individually remain trapped for a longer time; or 2) by increasing the detuning from the lowest allowed dipole transition. An increased detuning can primarily be achieved by replacing the trapping laser with a source that can produce light at a larger wavelength. The closest realization of a conservative optical trap can be achieved using a $\text{CO}_2$ laser to produce light at a $10.6\,\text{μm}$ wavelength. Such a quasi-electrostatic trap can reach scattering rates down to a $10\,\text{mHz}$ regime, but it may also cause mixing of the $A^2\Pi$ states, which would be disfavorable for the loading scheme. It may also be possible to suppress certain electronic transitions by appropriately choosing the trap polarization. For instance, in $\text{BaF}$ the lowest dipole allowed transition $X^2\Sigma\to A^2\Pi$ can be suppressed by using light that is polarized parallel to the quantization axis.\\
\indent An additional factor that limits the lifetime of the trapped molecules originates from the coupling of blackbody radiation to the rovibrational transitions of the molecule, as they may alter how the molecule interacts with the trap. Such rovibrational lifetimes for $\text{BaF}$ were previously calculated to be $1.8\,\text{s}$ at room temperature~\cite{buhmann_2008}, and can be significantly improved by cooling down the surroundings of the molecule.
\section{Conclusion}
\indent In conclusion, we have outlined the main considerations needed to realize a single-photon loading scheme. The scheme makes use of the state-dependent potential in an inhomogeneous electric field. Using $\text{BaF}$ as a prototype molecule, we have estimated the loading efficiency to be of the order $\sim0.04\%$ for a typical Stark decelerated beam, which comes down to $\sim750\,\text{molecules}$ per shot. The velocity-selective contribution to this efficiency can be improved to about $0.52\%$, or $10^4$ molecules, when synchronously modulating the pump laser frequency with the arrival time of different velocity components in a pulsed molecular beam. An additional factor 2 can potentially be gained when also modulating the electric field strength. The irreversibility of the loading process allows the accumulation of larger numbers of molecules over time, where the maximum accumulation number is limited by the trap lifetime. This single-photon loading scheme does not rely on optical cycling, which is beneficial for molecules that are difficult to laser cool. The ability to efficiently trap complex molecular species opens new opportunities to study cold molecular chemistry, quantum information, and fundamental interactions using optical traps.
\begin{acknowledgments}
This research was made possible through funding from the Dutch Research Council (NWO) through grants XL21.074 and VI.C.212.016.
\end{acknowledgments}
\bibliography{biblio.bib}
\end{document}

%% file: scheme.pdf_tex
\begingroup%
  \makeatletter%
  \providecommand\color[2][]{%
    \errmessage{(Inkscape) Color is used for the text in Inkscape, but the package 'color.sty' is not loaded}%
    \renewcommand\color[2][]{}%
  }%
  \providecommand\transparent[1]{%
    \errmessage{(Inkscape) Transparency is used (non-zero) for the text in Inkscape, but the package 'transparent.sty' is not loaded}%
    \renewcommand\transparent[1]{}%
  }%
  \providecommand\rotatebox[2]{#2}%
  \newcommand*\fsize{\dimexpr\f@size pt\relax}%
  \newcommand*\lineheight[1]{\fontsize{\fsize}{#1\fsize}\selectfont}%
  \ifx\svgwidth\undefined%
    \setlength{\unitlength}{504bp}%
    \ifx\svgscale\undefined%
      \relax%
    \else%
      \setlength{\unitlength}{\unitlength * \real{\svgscale}}%
    \fi%
  \else%
    \setlength{\unitlength}{\svgwidth}%
  \fi%
  \global\let\svgwidth\undefined%
  \global\let\svgscale\undefined%
  \makeatother%
  \begin{picture}(1,0.71428571)%
    \lineheight{1}%
    \setlength\tabcolsep{0pt}%
    \put(0,0){\includegraphics[width=\unitlength,page=1]{scheme.pdf}}%
    \put(0.64595779,0.03881635){\makebox(0,0)[lt]{\lineheight{1.25}\smash{\begin{tabular}[t]{l}{\normalfont\fontsize{21.599999999999998}{21.599999999999998}\selectfont{$z=0$}}\end{tabular}}}}%
    \put(0,0){\includegraphics[width=\unitlength,page=2]{scheme.pdf}}%
    \put(0.73737302,0.59295361){\makebox(0,0)[lt]{\lineheight{1.25}\smash{\begin{tabular}[t]{l}{\normalfont\fontsize{19.2}{19.2}\selectfont{A$^2\Pi_{1/2}$}}\end{tabular}}}}%
    \put(0.57109524,0.34968465){\makebox(0,0)[lt]{\lineheight{1.25}\smash{\begin{tabular}[t]{l}{\normalfont\fontsize{31.2}{31.2}\selectfont{$\ket{t}$}}\end{tabular}}}}%
    \put(0.55923016,0.58125222){\makebox(0,0)[lt]{\lineheight{1.25}\smash{\begin{tabular}[t]{l}{\normalfont\fontsize{31.2}{31.2}\selectfont{$\ket{e}$}}\end{tabular}}}}%
    \put(0.75287302,0.21356572){\makebox(0,0)[lt]{\lineheight{1.25}\smash{\begin{tabular}[t]{l}{\normalfont\fontsize{19.2}{19.2}\selectfont{$N=0$}}\end{tabular}}}}%
    \put(0.17394841,0.18405754){\makebox(0,0)[lt]{\lineheight{1.25}\smash{\begin{tabular}[t]{l}{\normalfont\fontsize{31.2}{31.2}\selectfont{$\ket{i}$}}\end{tabular}}}}%
    \put(0.75287302,0.36533863){\makebox(0,0)[lt]{\lineheight{1.25}\smash{\begin{tabular}[t]{l}{\normalfont\fontsize{18.0}{18.0}\selectfont{ $N=2$}}\end{tabular}}}}%
    \put(0.75287302,0.33513318){\makebox(0,0)[lt]{\lineheight{1.25}\smash{\begin{tabular}[t]{l}{\normalfont\fontsize{15.6}{15.6}\selectfont{$m_N=\pm1$}}\end{tabular}}}}%
    \put(0.59999999,0.10552579){\color[rgb]{0.50196078,0.50196078,0.50196078}\makebox(0,0)[t]{\lineheight{1.25}\smash{\begin{tabular}[t]{c}{\normalfont\fontsize{19.2}{19.2}\selectfont{Electric field gradient}}\end{tabular}}}}%
    \put(0.17920387,0.1233001){\color[rgb]{0.50196078,0.50196078,0.50196078}\makebox(0,0)[lt]{\lineheight{1.25}\smash{\begin{tabular}[t]{l}{\normalfont\fontsize{19.2}{19.2}\selectfont{Bias}}\end{tabular}}}}%
    \put(0.17824901,0.08775149){\color[rgb]{0.50196078,0.50196078,0.50196078}\makebox(0,0)[lt]{\lineheight{1.25}\smash{\begin{tabular}[t]{l}{\normalfont\fontsize{19.2}{19.2}\selectfont{field}}\end{tabular}}}}%
    \put(0.1875,0.58663628){\makebox(0,0)[lt]{\lineheight{1.25}\smash{\begin{tabular}[t]{l}{\normalfont\fontsize{25.2}{25.2}\selectfont{$\hat{x}$}}\end{tabular}}}}%
    \put(0.23895833,0.58663628){\makebox(0,0)[lt]{\lineheight{1.25}\smash{\begin{tabular}[t]{l}{\normalfont\fontsize{25.2}{25.2}\selectfont{$\hat{z}$}}\end{tabular}}}}%
    \put(0.07133463,0.1863258){\rotatebox{90}{\makebox(0,0)[lt]{\lineheight{1.25}\smash{\begin{tabular}[t]{l}{\normalfont\fontsize{25.2}{25.2}\selectfont{Potential Energy}}\end{tabular}}}}}%
  \end{picture}%
\endgroup%

%% file: energy.pdf_tex
\begingroup%
  \makeatletter%
  \providecommand\color[2][]{%
    \errmessage{(Inkscape) Color is used for the text in Inkscape, but the package 'color.sty' is not loaded}%
    \renewcommand\color[2][]{}%
  }%
  \providecommand\transparent[1]{%
    \errmessage{(Inkscape) Transparency is used (non-zero) for the text in Inkscape, but the package 'transparent.sty' is not loaded}%
    \renewcommand\transparent[1]{}%
  }%
  \providecommand\rotatebox[2]{#2}%
  \newcommand*\fsize{\dimexpr\f@size pt\relax}%
  \newcommand*\lineheight[1]{\fontsize{\fsize}{#1\fsize}\selectfont}%
  \ifx\svgwidth\undefined%
    \setlength{\unitlength}{648bp}%
    \ifx\svgscale\undefined%
      \relax%
    \else%
      \setlength{\unitlength}{\unitlength * \real{\svgscale}}%
    \fi%
  \else%
    \setlength{\unitlength}{\svgwidth}%
  \fi%
  \global\let\svgwidth\undefined%
  \global\let\svgscale\undefined%
  \makeatother%
  \begin{picture}(1,0.44444444)%
    \lineheight{1}%
    \setlength\tabcolsep{0pt}%
    \put(0,0){\includegraphics[width=\unitlength,page=1]{energy.pdf}}%
    \put(0.11709877,0.01465231){\makebox(0,0)[lt]{\lineheight{1.25}\smash{\begin{tabular}[t]{l}{\normalfont\fontsize{25.2}{25.2}\selectfont{$0$}}\end{tabular}}}}%
    \put(0,0){\includegraphics[width=\unitlength,page=2]{energy.pdf}}%
    \put(0.27009719,0.01465231){\makebox(0,0)[lt]{\lineheight{1.25}\smash{\begin{tabular}[t]{l}{\normalfont\fontsize{25.2}{25.2}\selectfont{$20$}}\end{tabular}}}}%
    \put(0,0){\includegraphics[width=\unitlength,page=3]{energy.pdf}}%
    \put(0.43099684,0.01465231){\makebox(0,0)[lt]{\lineheight{1.25}\smash{\begin{tabular}[t]{l}{\normalfont\fontsize{25.2}{25.2}\selectfont{$40$}}\end{tabular}}}}%
    \put(0,0){\includegraphics[width=\unitlength,page=4]{energy.pdf}}%
    \put(0.59189649,0.01465231){\makebox(0,0)[lt]{\lineheight{1.25}\smash{\begin{tabular}[t]{l}{\normalfont\fontsize{25.2}{25.2}\selectfont{$60$}}\end{tabular}}}}%
    \put(0.36634949,-0.03073228){\makebox(0,0)[t]{\lineheight{1.25}\smash{\begin{tabular}[t]{c}{\normalfont\fontsize{25.2}{25.2}\selectfont{Electric field (kV/cm)}}\end{tabular}}}}%
    \put(0,0){\includegraphics[width=\unitlength,page=5]{energy.pdf}}%
    \put(0.04994479,0.039701){\makebox(0,0)[lt]{\lineheight{1.25}\smash{\begin{tabular}[t]{l}{\normalfont\fontsize{25.2}{25.2}\selectfont{$-20$}}\end{tabular}}}}%
    \put(0,0){\includegraphics[width=\unitlength,page=6]{energy.pdf}}%
    \put(0.09414232,0.12525656){\makebox(0,0)[lt]{\lineheight{1.25}\smash{\begin{tabular}[t]{l}{\normalfont\fontsize{25.2}{25.2}\selectfont{$0$}}\end{tabular}}}}%
    \put(0,0){\includegraphics[width=\unitlength,page=7]{energy.pdf}}%
    \put(0.07833985,0.21081211){\makebox(0,0)[lt]{\lineheight{1.25}\smash{\begin{tabular}[t]{l}{\normalfont\fontsize{25.2}{25.2}\selectfont{$20$}}\end{tabular}}}}%
    \put(0,0){\includegraphics[width=\unitlength,page=8]{energy.pdf}}%
    \put(0.07833985,0.29636767){\makebox(0,0)[lt]{\lineheight{1.25}\smash{\begin{tabular}[t]{l}{\normalfont\fontsize{25.2}{25.2}\selectfont{$40$}}\end{tabular}}}}%
    \put(0,0){\includegraphics[width=\unitlength,page=9]{energy.pdf}}%
    \put(0.07833985,0.38192323){\makebox(0,0)[lt]{\lineheight{1.25}\smash{\begin{tabular}[t]{l}{\normalfont\fontsize{25.2}{25.2}\selectfont{$60$}}\end{tabular}}}}%
    \put(0,0){\includegraphics[width=\unitlength,page=10]{energy.pdf}}%
    \put(0.13304498,0.14184973){\makebox(0,0)[lt]{\lineheight{1.25}\smash{\begin{tabular}[t]{l}{\normalfont\fontsize{21.599999999999998}{21.599999999999998}\selectfont{N$=0$}}\end{tabular}}}}%
    \put(0.17491371,0.07383584){\makebox(0,0)[lt]{\lineheight{1.25}\smash{\begin{tabular}[t]{l}{\normalfont\fontsize{19.2}{19.2}\selectfont{$|0;0\rangle$}}\end{tabular}}}}%
    \put(0.13304498,0.20637809){\makebox(0,0)[lt]{\lineheight{1.25}\smash{\begin{tabular}[t]{l}{\normalfont\fontsize{21.599999999999998}{21.599999999999998}\selectfont{N$=1$}}\end{tabular}}}}%
    \put(0.25742001,0.11661362){\makebox(0,0)[lt]{\lineheight{1.25}\smash{\begin{tabular}[t]{l}{\normalfont\fontsize{19.2}{19.2}\selectfont{$\ket{1;\pm1}$}}\end{tabular}}}}%
    \put(0.33581336,0.17436362){\makebox(0,0)[lt]{\lineheight{1.25}\smash{\begin{tabular}[t]{l}{\normalfont\fontsize{19.2}{19.2}\selectfont{$\ket{1;0}$}}\end{tabular}}}}%
    \put(0.13304498,0.31509973){\makebox(0,0)[lt]{\lineheight{1.25}\smash{\begin{tabular}[t]{l}{\normalfont\fontsize{21.599999999999998}{21.599999999999998}\selectfont{N$=2$}}\end{tabular}}}}%
    \put(0.24133004,0.25350251){\makebox(0,0)[lt]{\lineheight{1.25}\smash{\begin{tabular}[t]{l}{\normalfont\fontsize{19.2}{19.2}\selectfont{$\ket{2;\pm2}$}}\end{tabular}}}}%
    \put(0.37004977,0.27703029){\makebox(0,0)[lt]{\lineheight{1.25}\smash{\begin{tabular}[t]{l}{\normalfont\fontsize{19.2}{19.2}\selectfont{$\ket{2;\pm1}$}}\end{tabular}}}}%
    \put(0.44844312,0.3326414){\makebox(0,0)[lt]{\lineheight{1.25}\smash{\begin{tabular}[t]{l}{\normalfont\fontsize{19.2}{19.2}\selectfont{$\ket{2;0}$}}\end{tabular}}}}%
    \put(0.125,0.42533333){\makebox(0,0)[lt]{\lineheight{1.25}\smash{\begin{tabular}[t]{l}{\normalfont\fontsize{25.2}{25.2}\selectfont{(GHz)}}\end{tabular}}}}%
    \put(0.04776817,0.44928889){\makebox(0,0)[lt]{\lineheight{1.25}\smash{\begin{tabular}[t]{l}{\normalfont\fontsize{25.2}{25.2}\selectfont{(a)}}\end{tabular}}}}%
    \put(0,0){\includegraphics[width=\unitlength,page=11]{energy.pdf}}%
    \put(0.7439967,0.05389221){\makebox(0,0)[rt]{\lineheight{1.25}\smash{\begin{tabular}[t]{r}{\normalfont\fontsize{25.2}{25.2}\selectfont{0}}\end{tabular}}}}%
    \put(0,0){\includegraphics[width=\unitlength,page=12]{energy.pdf}}%
    \put(0.7439967,0.13585173){\makebox(0,0)[rt]{\lineheight{1.25}\smash{\begin{tabular}[t]{r}{\normalfont\fontsize{25.2}{25.2}\selectfont{4000}}\end{tabular}}}}%
    \put(0,0){\includegraphics[width=\unitlength,page=13]{energy.pdf}}%
    \put(0.7439967,0.21781125){\makebox(0,0)[rt]{\lineheight{1.25}\smash{\begin{tabular}[t]{r}{\normalfont\fontsize{25.2}{25.2}\selectfont{8000}}\end{tabular}}}}%
    \put(0,0){\includegraphics[width=\unitlength,page=14]{energy.pdf}}%
    \put(0.7439967,0.29977079){\makebox(0,0)[rt]{\lineheight{1.25}\smash{\begin{tabular}[t]{r}{\normalfont\fontsize{25.2}{25.2}\selectfont{12000}}\end{tabular}}}}%
    \put(0,0){\includegraphics[width=\unitlength,page=15]{energy.pdf}}%
    \put(0.7439967,0.38173033){\makebox(0,0)[rt]{\lineheight{1.25}\smash{\begin{tabular}[t]{r}{\normalfont\fontsize{25.2}{25.2}\selectfont{16000}}\end{tabular}}}}%
    \put(0,0){\includegraphics[width=\unitlength,page=16]{energy.pdf}}%
    \put(0.91119366,0.33873996){\makebox(0,0)[lt]{\lineheight{1.25}\smash{\begin{tabular}[t]{l}{\normalfont\fontsize{25.2}{25.2}\selectfont{712}}\end{tabular}}}}%
    \put(0,0){\includegraphics[width=\unitlength,page=17]{energy.pdf}}%
    \put(0.91119366,0.28921516){\makebox(0,0)[lt]{\lineheight{1.25}\smash{\begin{tabular}[t]{l}{\normalfont\fontsize{25.2}{25.2}\selectfont{860}}\end{tabular}}}}%
    \put(0,0){\includegraphics[width=\unitlength,page=18]{energy.pdf}}%
    \put(0.91119366,0.24353479){\makebox(0,0)[lt]{\lineheight{1.25}\smash{\begin{tabular}[t]{l}{\normalfont\fontsize{25.2}{25.2}\selectfont{1064}}\end{tabular}}}}%
    \put(0,0){\includegraphics[width=\unitlength,page=19]{energy.pdf}}%
    \put(0.79009516,0.05677749){\makebox(0,0)[lt]{\lineheight{1.25}\smash{\begin{tabular}[t]{l}{\normalfont\fontsize{21.599999999999998}{21.599999999999998}\selectfont{X$^2\Sigma^+$}}\end{tabular}}}}%
    \put(0.79801182,0.27025928){\makebox(0,0)[lt]{\lineheight{1.25}\smash{\begin{tabular}[t]{l}{\normalfont\fontsize{21.599999999999998}{21.599999999999998}\selectfont{A'$^2\Delta$}}\end{tabular}}}}%
    \put(0.80092849,0.30327873){\makebox(0,0)[lt]{\lineheight{1.25}\smash{\begin{tabular}[t]{l}{\normalfont\fontsize{21.599999999999998}{21.599999999999998}\selectfont{A$^2\Pi$}}\end{tabular}}}}%
    \put(0.79009516,0.34445881){\makebox(0,0)[lt]{\lineheight{1.25}\smash{\begin{tabular}[t]{l}{\normalfont\fontsize{21.599999999999998}{21.599999999999998}\selectfont{B$^2\Sigma^+$}}\end{tabular}}}}%
    \put(0.67461521,0.42570708){\makebox(0,0)[lt]{\lineheight{1.25}\smash{\begin{tabular}[t]{l}{\normalfont\fontsize{25.2}{25.2}\selectfont{(cm$^{-1}$)}}\end{tabular}}}}%
    \put(0.87103808,0.42533333){\makebox(0,0)[lt]{\lineheight{1.25}\smash{\begin{tabular}[t]{l}{\normalfont\fontsize{25.2}{25.2}\selectfont{(nm)}}\end{tabular}}}}%
    \put(0.61038062,0.44928889){\makebox(0,0)[lt]{\lineheight{1.25}\smash{\begin{tabular}[t]{l}{\normalfont\fontsize{25.2}{25.2}\selectfont{(b)}}\end{tabular}}}}%
  \end{picture}%
\endgroup%

%% file: acceptance.pdf_tex
\begingroup%
  \makeatletter%
  \providecommand\color[2][]{%
    \errmessage{(Inkscape) Color is used for the text in Inkscape, but the package 'color.sty' is not loaded}%
    \renewcommand\color[2][]{}%
  }%
  \providecommand\transparent[1]{%
    \errmessage{(Inkscape) Transparency is used (non-zero) for the text in Inkscape, but the package 'transparent.sty' is not loaded}%
    \renewcommand\transparent[1]{}%
  }%
  \providecommand\rotatebox[2]{#2}%
  \newcommand*\fsize{\dimexpr\f@size pt\relax}%
  \newcommand*\lineheight[1]{\fontsize{\fsize}{#1\fsize}\selectfont}%
  \ifx\svgwidth\undefined%
    \setlength{\unitlength}{1296bp}%
    \ifx\svgscale\undefined%
      \relax%
    \else%
      \setlength{\unitlength}{\unitlength * \real{\svgscale}}%
    \fi%
  \else%
    \setlength{\unitlength}{\svgwidth}%
  \fi%
  \global\let\svgwidth\undefined%
  \global\let\svgscale\undefined%
  \makeatother%
  \begin{picture}(1,0.27777778)%
    \lineheight{1}%
    \setlength\tabcolsep{0pt}%
    \put(0,0){\includegraphics[width=\unitlength,page=1]{acceptance.pdf}}%
    \put(-0.00540123,0.15257402){\makebox(0,0)[rt]{\lineheight{1.25}\smash{\begin{tabular}[t]{r}{\normalfont\fontsize{25.2}{25.2}\selectfont{10}}\end{tabular}}}}%
    \put(0,0){\includegraphics[width=\unitlength,page=2]{acceptance.pdf}}%
    \put(-0.00540123,0.20019306){\makebox(0,0)[rt]{\lineheight{1.25}\smash{\begin{tabular}[t]{r}{\normalfont\fontsize{25.2}{25.2}\selectfont{10.1}}\end{tabular}}}}%
    \put(0,0){\includegraphics[width=\unitlength,page=3]{acceptance.pdf}}%
    \put(-0.00540123,0.24781211){\makebox(0,0)[rt]{\lineheight{1.25}\smash{\begin{tabular}[t]{r}{\normalfont\fontsize{25.2}{25.2}\selectfont{10.2}}\end{tabular}}}}%
    \put(-0.08458796,0.14635851){\rotatebox{90}{\makebox(0,0)[lt]{\lineheight{1.25}\smash{\begin{tabular}[t]{l}{\normalfont\fontsize{28.799999999999997}{28.799999999999997}\selectfont{Initial velocity}}\end{tabular}}}}}%
    \put(-0.06385127,0.18692419){\rotatebox{90}{\makebox(0,0)[lt]{\lineheight{1.25}\smash{\begin{tabular}[t]{l}{\normalfont\fontsize{28.799999999999997}{28.799999999999997}\selectfont{(m/s)}}\end{tabular}}}}}%
    \put(0,0){\includegraphics[width=\unitlength,page=4]{acceptance.pdf}}%
    \put(0.21608824,0.18981354){\makebox(0,0)[lt]{\lineheight{1.25}\smash{\begin{tabular}[t]{l}{\normalfont\fontsize{25.2}{25.2}\selectfont{0.04 m/s}}\end{tabular}}}}%
    \put(0.015,0.25828414){\makebox(0,0)[lt]{\lineheight{1.25}\smash{\begin{tabular}[t]{l}{\normalfont\fontsize{28.799999999999997}{28.799999999999997}\selectfont{$\ket{i}$}}\end{tabular}}}}%
    \put(0,0){\includegraphics[width=\unitlength,page=5]{acceptance.pdf}}%
    \put(0.03066108,-0.01740005){\makebox(0,0)[lt]{\lineheight{1.25}\smash{\begin{tabular}[t]{l}{\normalfont\fontsize{25.2}{25.2}\selectfont{$-100$}}\end{tabular}}}}%
    \put(0,0){\includegraphics[width=\unitlength,page=6]{acceptance.pdf}}%
    \put(0.13599129,-0.01740005){\makebox(0,0)[lt]{\lineheight{1.25}\smash{\begin{tabular}[t]{l}{\normalfont\fontsize{25.2}{25.2}\selectfont{$0$}}\end{tabular}}}}%
    \put(0,0){\includegraphics[width=\unitlength,page=7]{acceptance.pdf}}%
    \put(0.21393723,-0.01740005){\makebox(0,0)[lt]{\lineheight{1.25}\smash{\begin{tabular}[t]{l}{\normalfont\fontsize{25.2}{25.2}\selectfont{$100$}}\end{tabular}}}}%
    \put(0.15,-0.03824097){\makebox(0,0)[t]{\lineheight{1.25}\smash{\begin{tabular}[t]{c}{\normalfont\fontsize{28.799999999999997}{28.799999999999997}\selectfont{Position z ($\mu$m)}}\end{tabular}}}}%
    \put(0,0){\includegraphics[width=\unitlength,page=8]{acceptance.pdf}}%
    \put(-0.04996142,0.01637955){\makebox(0,0)[lt]{\lineheight{1.25}\smash{\begin{tabular}[t]{l}{\normalfont\fontsize{25.2}{25.2}\selectfont{$-300$}}\end{tabular}}}}%
    \put(0,0){\includegraphics[width=\unitlength,page=9]{acceptance.pdf}}%
    \put(-0.0157716,0.04971288){\makebox(0,0)[lt]{\lineheight{1.25}\smash{\begin{tabular}[t]{l}{\normalfont\fontsize{25.2}{25.2}\selectfont{$0$}}\end{tabular}}}}%
    \put(0,0){\includegraphics[width=\unitlength,page=10]{acceptance.pdf}}%
    \put(-0.03635031,0.08304621){\makebox(0,0)[lt]{\lineheight{1.25}\smash{\begin{tabular}[t]{l}{\normalfont\fontsize{25.2}{25.2}\selectfont{$300$}}\end{tabular}}}}%
    \put(-0.08458796,-0.0299783){\rotatebox{90}{\makebox(0,0)[lt]{\lineheight{1.25}\smash{\begin{tabular}[t]{l}{\normalfont\fontsize{28.799999999999997}{28.799999999999997}\selectfont{Transition energy}}\end{tabular}}}}}%
    \put(-0.06385127,0.02296296){\rotatebox{90}{\makebox(0,0)[lt]{\lineheight{1.25}\smash{\begin{tabular}[t]{l}{\normalfont\fontsize{28.799999999999997}{28.799999999999997}\selectfont{(MHz)}}\end{tabular}}}}}%
    \put(0,0){\includegraphics[width=\unitlength,page=11]{acceptance.pdf}}%
    \put(0.216,0.0516708){\makebox(0,0)[lt]{\lineheight{1.25}\smash{\begin{tabular}[t]{l}{\normalfont\fontsize{25.2}{25.2}\selectfont{50 MHz}}\end{tabular}}}}%
    \put(0.045,0.01161748){\makebox(0,0)[lt]{\lineheight{1.25}\smash{\begin{tabular}[t]{l}{\normalfont\fontsize{28.799999999999997}{28.799999999999997}\selectfont{$\ket{i}\to\ket{e}$}}\end{tabular}}}}%
    \put(-0.006,0.1){\makebox(0,0)[rt]{\lineheight{1.25}\smash{\begin{tabular}[t]{r}{\normalfont\fontsize{19.2}{19.2}\selectfont{+3.49e8}}\end{tabular}}}}%
    \put(0,0){\includegraphics[width=\unitlength,page=12]{acceptance.pdf}}%
    \put(0.43325926,0.25792843){\makebox(0,0)[lt]{\lineheight{1.25}\smash{\begin{tabular}[t]{l}{\normalfont\fontsize{38.4}{38.4}\selectfont{$t_{1}$}}\end{tabular}}}}%
    \put(0,0){\includegraphics[width=\unitlength,page=13]{acceptance.pdf}}%
    \put(0.46243201,0.00702498){\makebox(0,0)[t]{\lineheight{1.25}\smash{\begin{tabular}[t]{c}{\normalfont\fontsize{33.6}{33.6}\selectfont{0}}\end{tabular}}}}%
    \put(0,0){\includegraphics[width=\unitlength,page=14]{acceptance.pdf}}%
    \put(0.3952194,0.07079996){\makebox(0,0)[rt]{\lineheight{1.25}\smash{\begin{tabular}[t]{r}{\normalfont\fontsize{33.6}{33.6}\selectfont{1}}\end{tabular}}}}%
    \put(0,0){\includegraphics[width=\unitlength,page=15]{acceptance.pdf}}%
    \put(0.3952194,0.02179181){\makebox(0,0)[rt]{\lineheight{1.25}\smash{\begin{tabular}[t]{r}{\normalfont\fontsize{33.6}{33.6}\selectfont{0}}\end{tabular}}}}%
    \put(0,0){\includegraphics[width=\unitlength,page=16]{acceptance.pdf}}%
    \put(0.52925926,0.25792843){\makebox(0,0)[lt]{\lineheight{1.25}\smash{\begin{tabular}[t]{l}{\normalfont\fontsize{38.4}{38.4}\selectfont{$t_{2}$}}\end{tabular}}}}%
    \put(0,0){\includegraphics[width=\unitlength,page=17]{acceptance.pdf}}%
    \put(0.55843198,0.00702498){\makebox(0,0)[t]{\lineheight{1.25}\smash{\begin{tabular}[t]{c}{\normalfont\fontsize{33.6}{33.6}\selectfont{0}}\end{tabular}}}}%
    \put(0,0){\includegraphics[width=\unitlength,page=18]{acceptance.pdf}}%
    \put(0.62525926,0.25792843){\makebox(0,0)[lt]{\lineheight{1.25}\smash{\begin{tabular}[t]{l}{\normalfont\fontsize{38.4}{38.4}\selectfont{$t_{3}$}}\end{tabular}}}}%
    \put(0,0){\includegraphics[width=\unitlength,page=19]{acceptance.pdf}}%
    \put(0.65443199,0.00702498){\makebox(0,0)[t]{\lineheight{1.25}\smash{\begin{tabular}[t]{c}{\normalfont\fontsize{33.6}{33.6}\selectfont{0}}\end{tabular}}}}%
    \put(0,0){\includegraphics[width=\unitlength,page=20]{acceptance.pdf}}%
    \put(0.72125926,0.25792843){\makebox(0,0)[lt]{\lineheight{1.25}\smash{\begin{tabular}[t]{l}{\normalfont\fontsize{38.4}{38.4}\selectfont{$t_{4}$}}\end{tabular}}}}%
    \put(0,0){\includegraphics[width=\unitlength,page=21]{acceptance.pdf}}%
    \put(0.750432,0.00702498){\makebox(0,0)[t]{\lineheight{1.25}\smash{\begin{tabular}[t]{c}{\normalfont\fontsize{33.6}{33.6}\selectfont{0}}\end{tabular}}}}%
    \put(0,0){\includegraphics[width=\unitlength,page=22]{acceptance.pdf}}%
    \put(0.81725926,0.25792843){\makebox(0,0)[lt]{\lineheight{1.25}\smash{\begin{tabular}[t]{l}{\normalfont\fontsize{38.4}{38.4}\selectfont{$t_{5}$}}\end{tabular}}}}%
    \put(0,0){\includegraphics[width=\unitlength,page=23]{acceptance.pdf}}%
    \put(0.84643197,0.00702498){\makebox(0,0)[t]{\lineheight{1.25}\smash{\begin{tabular}[t]{c}{\normalfont\fontsize{33.6}{33.6}\selectfont{0}}\end{tabular}}}}%
    \put(0,0){\includegraphics[width=\unitlength,page=24]{acceptance.pdf}}%
    \put(0.89408,0.06081395){\makebox(0,0)[lt]{\lineheight{1.25}\smash{\begin{tabular}[t]{l}{\normalfont\fontsize{28.799999999999997}{28.799999999999997}\selectfont{Electric field}}\end{tabular}}}}%
    \put(0.89408,0.04068007){\makebox(0,0)[lt]{\lineheight{1.25}\smash{\begin{tabular}[t]{l}{\normalfont\fontsize{28.799999999999997}{28.799999999999997}\selectfont{($1/E_{max}$)}}\end{tabular}}}}%
    \put(0.89408,0.18548061){\makebox(0,0)[lt]{\lineheight{1.25}\smash{\begin{tabular}[t]{l}{\normalfont\fontsize{28.799999999999997}{28.799999999999997}\selectfont{Potential}}\end{tabular}}}}%
    \put(0.89408,0.16474392){\makebox(0,0)[lt]{\lineheight{1.25}\smash{\begin{tabular}[t]{l}{\normalfont\fontsize{28.799999999999997}{28.799999999999997}\selectfont{energy}}\end{tabular}}}}%
    \put(0.59999998,-0.01407117){\makebox(0,0)[lt]{\lineheight{1.25}\smash{\begin{tabular}[t]{l}{\normalfont\fontsize{28.799999999999997}{28.799999999999997}\selectfont{Position z}}\end{tabular}}}}%
    \put(-0.05,0.29166667){\makebox(0,0)[rt]{\lineheight{1.25}\smash{\begin{tabular}[t]{r}{\normalfont\fontsize{28.799999999999997}{28.799999999999997}\selectfont{(a)}}\end{tabular}}}}%
    \put(0.37599999,0.29166667){\makebox(0,0)[rt]{\lineheight{1.25}\smash{\begin{tabular}[t]{r}{\normalfont\fontsize{28.799999999999997}{28.799999999999997}\selectfont{(b)}}\end{tabular}}}}%
  \end{picture}%
\endgroup%